\documentclass[conference]{IEEEtran}
\IEEEoverridecommandlockouts
\usepackage{cite}
\usepackage{amsmath}
\usepackage{amssymb}
\usepackage{multirow}
\usepackage{amsfonts}
\usepackage{algorithm}
\usepackage{algorithmic}
\usepackage{graphicx}
\usepackage{textcomp}
\usepackage{xcolor}
\usepackage{mathrsfs}
\usepackage{booktabs}  
\usepackage{rotating}
\usepackage{array}
\usepackage{siunitx}
\usepackage[marginal]{footmisc}

\newcolumntype{R}{>{\raggedleft\arraybackslash}p{1.2cm}}

\newcolumntype{C}{>{\centering\arraybackslash}p{1.8cm}}
\def\BibTeX{{\rm B\kern-.05em{\sc i\kern-.025em b}\kern-.08em
    T\kern-.1667em\lower.7ex\hbox{E}\kern-.125emX}}

\begin{document}

% DeclareMathAlphabet{mathcal}{OMS}{cmsy}{m}{n}
% \title{Conference Paper Title*\\
% {\footnotesize \textsuperscript{*}Note: Sub-titles are not captured for https://ieeexplore.ieee.org  and
\title{A Simple Contrastive Framework Of Item Tokenization For Generative Recommendation}

\author{
    \IEEEauthorblockN{Penglong Zhai$^{1*}$, Yifang Yuan$^{1*}$, Fanyi Di$^{1*}$, Jie Li$^{1}$}
    \IEEEauthorblockN{Yue Liu$^{1}$, Chen Li$^{1}$, Jie Huang$^{1}$, Sicong Wang$^{1}$, Yao Xu$^{1}$, Xin Li$^{1\dag}$}
    \IEEEauthorblockA{$^1$ AMAP, Alibaba Group}
    % \IEEEauthorblockA{$^b$ Department of Computer Science and Technology, Tsinghua University, Beijing, China}
    \IEEEauthorblockA{\{zhaipenglong.zpl, yuanyifang.yyf, difanyi.dfy, lj313796\}@alibaba-inc.com}
    \IEEEauthorblockA{\{ly355576, lichen298097, jielu.hj, wsc488938, xuenuo.xy, beilai.bl\}@alibaba-inc.com}
}

\maketitle
% \footnotetext[1]{The two authors contribute equally to this work.}
\footnote{------------------------- }
\footnote{$^*$ Equal contribution.}
\footnote{$^\dag$ Corresponding author.}

\begin{abstract}
Generative retrieval-based recommendation has emerged as a promising paradigm aiming at directly generating the identifiers of the target candidates. However, in large-scale recommendation systems, this approach becomes increasingly cumbersome due to the redundancy and sheer scale of the token space. To overcome these limitations, recent research has explored the use of semantic tokens as an alternative to ID tokens, which typically leveraged reconstruction-based strategies, like RQ-VAE, to quantize content embeddings and significantly reduce the embedding size. However, reconstructive quantization aims for the precise reconstruction of each item embedding independently, which conflicts with the goal of generative retrieval tasks focusing more on differentiating among items. 
Moreover, multi-modal side information of items, such as descriptive text and images, geographical knowledge in location-based recommendation services, has been shown to be effective in improving recommendations by providing richer contexts for interactions. Nevertheless, effectively integrating such complementary knowledge into existing generative recommendation frameworks remains challenging. To overcome these challenges, we propose a novel unsupervised deep quantization exclusively based on contrastive learning, named SimCIT (a Simple Contrastive Item Tokenization framework). Specifically, different from existing reconstruction-based strategies, SimCIT propose to use a learnable residual quantization module to align with the signals from different modalities of the items, which combines multi-modal knowledge alignment and semantic tokenization in a mutually beneficial contrastive learning framework.
Extensive experiments across public datasets and a large-scale industrial dataset from various domains demonstrate SimCIT's effectiveness in LLM-based generative recommendation.

\end{abstract}


\section{Introduction}
\begin{figure}
    \centering
    \includegraphics[width=\linewidth]{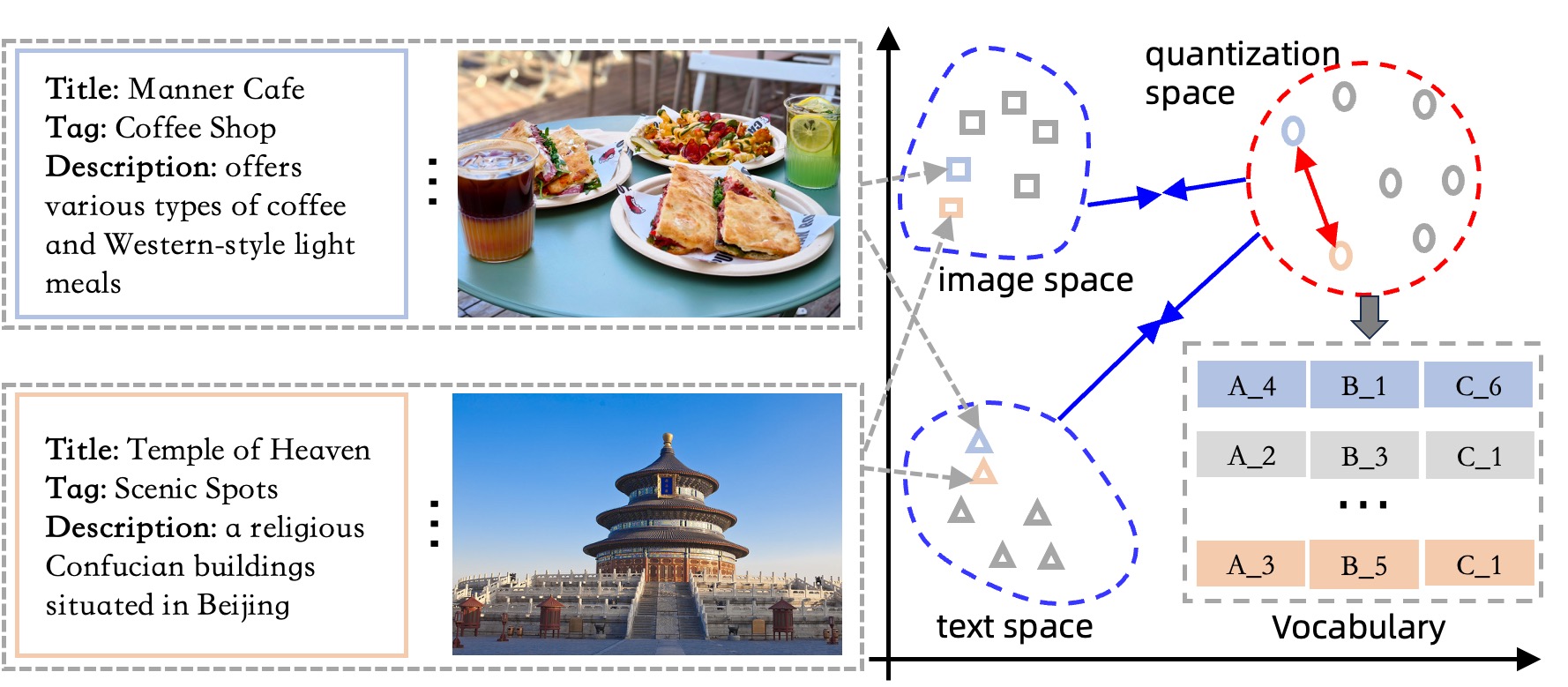}
    \caption{Illustration of the proposed SimCIT. We translate items from different domains and modalities into a new unified semantic ID, which can then serve as a bridge for transferring retrieval knowledge.}
    \label{fig:background}
\end{figure}

In the rapidly evolving landscape of digital services, recommendation systems have become pivotal in shaping content consumption habits \cite{youtube,gomez2015netflix}.
With the recent success of generative AI techniques, generative retrieval has emerged as a new retrieval paradigm for recommendation, redefining item retrieval as a sequence generation problem \cite{deldjoo2024review, Vaswani2017, Devlin2019, Sutskever2014, Rajput2023}.
TIGER \cite{Rajput2023} was the first to exemplify this approach with its two-stage model: semantic tokenization and autoregressive generation. 
As illustrated in Fig. \ref{fig:background}, in the tokenization, the model maps each item to a sequence of discrete semantic tokens or IDs, aiming to capture item multi-modal content semantics within these tokens. This ensures that semantically similar items are assigned similar tokens. In the generation stage, an sequence-to-sequence architecture is usually employed to encode historical token sequences and decode target item tokens autoregressively. 
During inference, the model leverages beam search coupled with token-to-item mapping to directly output top-$k$ candidate item IDs. This capability enables generative recommendation frameworks to achieve granular token-level user-item alignment while bypassing conventional non-differentiable ANN retrieval modules.

% However, in large-scale recommendation systems serving billions of users and items, this approach faces increasing challenges due to the redundancy and sheer scale of token spaces, necessitating more sophisticated representation learning techniques that can maintain both efficiency and effectiveness \cite{Rajput2023}. 
% Fig. 1. Illustration of our SimCIT. We translate items from different domains and modalities into a new unified semantic ID.

In this process, item tokenization \cite{10.1145/3640457.3688190,liu2024vector,Rajput2023,10.5555/3600270.3603090, Wang2024EAGERTG} acts as a key first step for training generative recommendation models. It simplifies the task by shrinking the vocabulary from millions of item IDs to just thousands of tokens, allowing faster decoding without requiring ANN search. 
To obtain high-quality semantic tokens, existing approaches typically utilize item textual descriptions paired with pretrained language models to generate semantic embeddings. These embeddings are then discretized into tokens through techniques like vector quantization \cite{liu2024vector, Rajput2023, 10.1145/3640457.3688190} or hierarchical k-means clustering \cite{10.5555/3600270.3603090, 10.5555/3600270.3601857, Wang2024EAGERTG}, corresponding to specific codes or clusters.
Among these techniques, RQ-VAE emerges as the predominant method for semantic tokenization, with its initially established through TIGER. 
These VAE-based methods \cite{Rajput2023,10.1145/3640457.3688178,10.1145/3627673.3679569} usually employ a residual quantization-based autoencoder to encode item's textual embeddings into hierarchical code sequences as identifiers, optimized through a reconstruction loss.
Reconstructive quantization aims for the precise reconstruction of each item embedding independently, which fundamentally conflicts with the nature of generative retrieval tasks aimed to identify the target items from the vocabulary.
This can lead to a suboptimal distribution of the learned semantic token space when applied to generative retrieval models. 
One example is that the common long-tail distribution observed in real-world item populations. Within high-frequency clusters, representation collapse leads the model to generate ambiguous token sequences for items with semantic proximity. This phenomenon highlights the need to capture latent topological relationships between items – an essential capability for maintaining recommendation systems' differentiation efficacy.

On the other hand, with the recent popularity of multimedia-centric scenarios like social media and micro-video websites, multi-modal recommendation systems are gaining widespread attention and adoption. 
% Moreover, modern recommendation scenarios increasingly involve rich multi-modal information that encompasses diverse data sources including textual descriptions, visual content, audio features, and contextual metadata. 
The information from different modalities characterizes diverse relationships among items. These modalities play a vital role in modeling item-item relationships and enhancing recommendation performance, especially for cold-start items.

For example, in POI (Point-of-Interest) recommendation tasks, 
% The importance of multi-modal contents becomes particularly evident in specialized domains such as location-based services (LBS), where Point-of-Interest (POI) recommendation systems must consider not only textual descriptions and visual imagery but also crucial 
spatial relationships and geographical knowledge have widely been proven critical for capturing intrinsic patterns of user mobility \cite{9133505,10.1145/3539618.3591770}.
These spatial dependencies, including proximity relationships and mobility patterns, are essential for accurate prediction but are often underutilized or oversimplified in previous methods of item tokenization.
Contemporary fusion paradigms employing primitive integration mechanisms (e.g. input-layer concatenation or prediction-stage score aggregation) fundamentally fail to capture the intricate interplay between multi-modal signals and their hierarchical semantic dependencies required for effective discriminative embedding learning. Such deficiencies critically impede industrial deployment where strict engineering constraints – particularly the competing demands of latency optimization, memory efficiency, and throughput maximization – exert measurable impacts on contractual service benchmarks and core business objectives.

To overcome the above limitations, we've introduced SimCIT, a scalable simple contrastive item tokenization framework that significantly boosts generative recommendation performance through the use of generated item identifier tokens which achieved by leveraging multi-modal information and contrastive learning. Unlike conventional VAE-based methods, SimCIT's framework eliminates the need for reconstruction loss, relying solely on contrastive learning. This approach not only preserves items' multi-modal features, such as text and image information, but also enhances the discrimination between different item tokens. Specifically, SimCIT employs a novel contrastive learning framework, treating multi-modal content of items (e.g. visual images and textual descriptions) as different ``views". It then applies a series of contrastive losses upon a soft residual quantization with these views to facilitate the fusion and alignment of identifiers. In this way, SimCIT implicitly aligns multi-modal features indirectly with an identifier. In other words, the identifier acts as a ``bridge" between modalities.

% In particular, 
% neighborhood relationships.
% Specifically, we enforce that reconstructed vectors are closer to their corresponding input vectors than to other vectors within a batch. This method thus preserves the neighborhood information
% between the input embedding and its reconstructed counterpart
% while enhancing dissimilarity from other items, facilitating clearer
% differentiation between each item and its neighbors. Our approach
% relaxes the requirement for exact reconstruction and instead prioritizes pairwise similarities between items. By leveraging contrastive loss, which maximizes the top-one probability within a batch, we more effectively capture the underlying distribution of item similarities. This provides a key advantage for recommendation tasks compared to RQ-VAE.
% It is end-to-end learnable together with the model parameters. To avoid abrupt changes during training, we further employ a soft search technique, which has been widely used in previous self-supervised learning works.
In summary, our contributions are highlighted as follows:
\begin{itemize}
    \item We propose, for the first time, a fully contrastive learning-based item tokenization framework for generative recommendation, named SimCIT, which is novel, simple, and scalable. Compared to previous reconstruction-based techniques, the proposed method achieves significant improvements in terms of discriminative capability and alleviating collision.
    % \item Building upon this scalable framework, we propose to learn the hierarchical identifiers by incorporating various multi-modal semantic information of the items, including text and image descriptions, collaborative signals, and geometric locations(for poi recommendation), which have been shown effective in traditional recommendation systems. 
    \item Building upon this scalable framework, we introduce a hierarchical identifier learning paradigm that systematically integrates heterogeneous item-side data modalities—spanning textual/image descriptions, collaborative filtering signals, and spatial graph (in POI recommendation tasks)—thereby extending the empirically validated feature integration strategies of conventional recommender systems to our tokenization architecture.
    \item We instantiate SimCIT on generative recommender models and conduct extensive experiments on various datasets from e-commerce and location-based recommendation, coupled with the in-depth investigation with diverse settings, demonstrating that SimCIT outperforms existing item tokenization methods for generative recommendation.
\end{itemize}

% The success of SimCIT in both rigorous academic evaluation and demanding industrial deployment underscores its practical value and theoretical soundness, establishing a new paradigm for semantic tokenization in generative recommendation systems. Our approach not only advances the state-of-the-art in recommendation algorithms through its novel contrastive learning framework but also provides a robust foundation for future research in multi-modal representation learning and large-scale AI systems, opening new avenues for cross-domain knowledge transfer and unified recommendation architectures.
The rest of the paper is organized as follows. We introduce related work to SimCIT in Section \ref{sec:related_work}, In Section \ref{sec:proposed_method}, we describe our framework, called Simple Contrastive Item Tokenzization. Experimental results are presented in Section \ref{sec:experiments}. Section \ref{sec:conclusion} concludes the paper.
% Distillation (CODIS), a
% The remainder of this paper presents our methodology in detail (Section II), describes comprehensive experimental evaluations across multiple domains (Section III), and discusses implications for both research advancement and industrial applications (Section IV). 

\section{Related work}
\label{sec:related_work}
\subsection{Generative Recommendation}
Traditional recommender systems typically adopt a two-stage retrieve-and-rank paradigm: first, retrieving a relevant subset from a vast candidate pool, and then precisely ranking this subset. However, this paradigm faces challenges related to retrieval efficiency, memory consumption, and cold-start issues \cite{Yang2024}. In recent years, generative recommender systems have emerged as a promising alternative, aiming to directly predict item identifiers (or their semantic encodings) rather than relying on a separate retrieval process \cite{Deng2025, Rajput2023}.
Among these works, Tay et al. \cite{10.5555/3600270.3601857}
introduced an end-to-end model that maps string queries directly to relevant document IDs, named DSI. 
% Specifically, 
% DSI models encode corpus information within their parameters, dramatically simplifying the retrieval process.
While DSI demonstrated potential in information retrieval by memorizing unidirectional mappings from pseudo-queries to document identifiers, Chen et al.\cite{chen2023understanding} indicated that it often struggles to distinguish relevant documents from random ones, thereby negatively impacting its retrieval effectiveness. 
% Moreover, DSI was not directly applied to the complexities of recommendation systems or the generation of sophisticated semantic IDs for items.
To adapt to the unique characteristics of recommendation tasks, several works have explored integrating the generative paradigm into recommender systems. Tiger \cite{Rajput2023} introduced one of the first semantic ID-based generative models for recommendation. It employs a Transformer-based sequence-to-sequence model that auto-regressively decodes semantic tuples of codewords, serving as Semantic IDs for items. This approach significantly outperformed existing state-of-the-art models and improved generalization for cold-start items. Following this, LIGER \cite{Yang2024} further investigates the unification of generative retrieval and sequential dense retrieval, which combines the strengths of both approaches.
% integrating sequential dense retrieval into generative retrieval to mitigate performance differences and enhance cold-start item recommendation. 
This work highlights the potential of combining different retrieval paradigms to overcome practical challenges in large-scale systems.
More recently, OneRec \cite{Deng2025} proposed a unified end-to-end generative recommendation framework, which utilized an encoder-decoder structure to encode user historical behavior sequences and then gradually decodes videos of user interest.
% It introduces a session-wise generation approach, which is more elegant and contextually coherent than traditional next-item prediction.
Furthermore, OneRec incorporates an Iterative Preference Alignment module \cite{Rafailov2023} to enhance the quality of generated results. These techniques achieved substantial improvements in industrial scenarios, demonstrating the significant promise of generative recommender systems.
% Generative recommender systems hold significant promise for addressing efficiency and memory challenges in large-scale recommendation by reframing the task as item identifier generation. They offer an alternative to traditional retrieve-and-rank pipelines.
% Despite these advances, existing generative recommendation methods still face significant challenges in the effective construction of semantic IDs, especially when dealing with multi-source heterogeneous data. Many approaches implicitly learn semantic IDs through reconstruction losses, which may not be optimal for discriminative recommendation tasks. Furthermore, there is a distinct lack of specialized designs for generating semantic IDs that can effectively encode and leverage the complex multi-modal and spatiotemporal characteristics inherent in POI data. The discriminative power and collision rate of generated IDs remain critical concerns for real-world applications.
% \subsection{Semantic ID Learning and Vector Quantization}
\subsection{Semantic Indexing and Tokenization}
Beyond the core generation framework, how to effectively index items has gained significant attention. While straightforward indexing methods like random or title-based indexing are simple, they just don't scale well for large, industrial-sized recommendation systems. To tackle this, recent research \cite{qu2024tokenreclearningtokenizeid, 10.1145/3627673.3679569, 10.1145/3640457.3688178, 10.1145/3640457.3688190, 10.5555/3600270.3601857} has gravitated towards semantic indexing techniques, which aim to categorize items based on their inherent content information. For instance, Hua et al.  \cite{10.1145/3624918.3625339}delved into item indexing methods specifically for LLM-based recommendation models, like P5\cite{f2f69092bf38447eaf9c02d9bd716456}. Within this area, two primary techniques stand out: vector quantization and hierarchical k-means clustering. Because of their discrete nature and resemblances to visual tokenization, this process is often called semantic tokenization. For example, TIGER\cite{Rajput2023} and LC-Rec\cite{zheng2024adaptinglargelanguagemodels} use residual quantization (RQ-VAE) on textual embeddings derived from item titles and descriptions for tokenization. On the other hand, Recforest\cite{10.5555/3600270.3603090} and EAGER\cite{Wang2024EAGERTG} employ hierarchical k-means clustering on item textual embeddings to create cluster indexes as tokens.

Furthermore, more recent studies, including EAGER, TokenRec\cite{qu2024tokenreclearningtokenizeid}, LETTER\cite{10.1145/3627673.3679569}, and CoST\cite{10.1145/3640457.3688178}, have been exploring how to integrate both semantic and collaborative information into the tokenization process. This marks a move towards more comprehensive and robust indexing methods that can capture both an item's inherent meaning and its relationships based on user interactions. However, this particular work zeroes in on a different aspect: how to leverage contrastive quantization to enhance semantic tokenization. By applying principles from contrastive learning, this approach seeks to generate more discriminative semantic tokens that better capture the unique characteristics of items, ultimately aiming for more precise and personalized item retrieval and recommendations within the system.

\subsection{Multi-modal Recommendation}
Multi-modal Representation Learning aims to learn effective representations that can capture information across different modalities, such as text, images, and graph structure knowledge. Transformer-based models \cite{pmlr-v139-radford21a} have shown remarkable success in multi-modal feature learning. These models typically learn transferable visual representations by leveraging corresponding natural language supervision. Models like FLAVA \cite{DBLP:conf/cvpr/SinghHGCGRK22} and Perceiver \cite{jaegle2021perceiver} have demonstrated the effectiveness of jointly pre-training transformers on unpaired images and text, while CLIP \cite{pmlr-v139-radford21a} has shown that contrastive objectives can effectively align representations from different modalities. Moreover, Fu et al. \cite{10.1145/3626772.3657725} propose a simple plug-and-play architecture using a Decoupled PEFT structure and exploiting both intra- and inter-modal adaptation.
% For more details, we refer readers to the survey (Xu et al.,
% 2023).

In recommender systems, multi-modal side information of items, such as descriptive text and images, has been shown to be effective in improving recommendations by providing richer contexts for interactions.
Recent works \cite{10.1145/3340531.3411947} propose to leverage various types of graph neural network (GNN) to fuse the multi-modal features. For example, Zhang et al. \cite{10.1145/3474085.3475259} designs a modality-aware layer to learn item-item structures for each modality and aggregates them to obtain latent item graphs. DualGNN \cite{9662655} proposes a multi-modal representation learning module to model the user attentions across modalities and inductively learn the user preference.

More recently, with the emergence of generative retrieval, some works propose to utilize the multi-modal contents of the items to learn the tokenization. Specifically, 
Zhai et al. \cite{zhai2025mmql} introduce quantitative translators to convert the
text and image content of items from various domains into a new and concise language, with all items sharing the same vocabulary. MMGRec \cite{Liu2024} utilizes a Graph RQ-VAE to construct item IDs from both multi-modal and collaborative information. 
Zhu et al. \cite{zhu2025unimodalboundariesgenerativerecommendation} reveal that GR models are particularly sensitive to different modalities
and examine the challenges in achieving effective generative retrieval when multiple modalities are available.

\subsection{Contrastive Representation Learning}
Contrastive learning has recently received increasing attention for it brings tremendous improvements on self-supervised representation learning, especially in the field of computer vision \cite{Chen2020SimCLR, He2020MoCo, Wang_Zeng_Chen_Dai_Xia_2022}.
It learns discriminative features by pulling representations of similar samples together and pulling apart those of dissimilar ones. 
Inspired by this paradigm, 
% Initially achieving breakthroughs in computer vision (e.g. SimCLR \cite{Chen2020SimCLR}, MoCo \cite{He2020MoCo}), the core idea of contrastive learning is to extract semantic information from data without manual annotations. 
% Subsequently, 
contrastive learning has been widely applied to various representation learning tasks, including recommendation systems \cite{Wu2021SGL, Zhou2021CLRec}, multi-modal fusion and alignment \cite{Liu2021,Xue2024,pmlr-v139-radford21a}.
Specifically, multi-modal contrastive learning has achieved remarkable success in aligning cross-modal representations, which typically takes the multi-modal information of the entities as different views.
For example, Radford et al. \cite{pmlr-v139-radford21a} propose to learn transferable visual representations with zero-shot strategy by training on a massive data set of image-text pairs through contrastive learning. 
One of the main challenge is feature collapsing, and in practice, a large number of negative samples are required, through either large batch size \cite{Chen2020SimCLR} or memory banks \cite{He2020MoCo} to alleviate this problem. Another problem is that this approach could consider only the stronger
modalities while ignoring the weaker ones. Liu et al.\cite{Liu2021} introduce a novel contrastive learning objective to addresses the issue of weaker modalities being ignored. It constructs new negative tuples using modalities that describe different scenes to contrast tuples, encouraging the model to examine correspondences between modalities within the same tuple.

\begin{figure*}
    \centering
    \includegraphics[width=\linewidth]{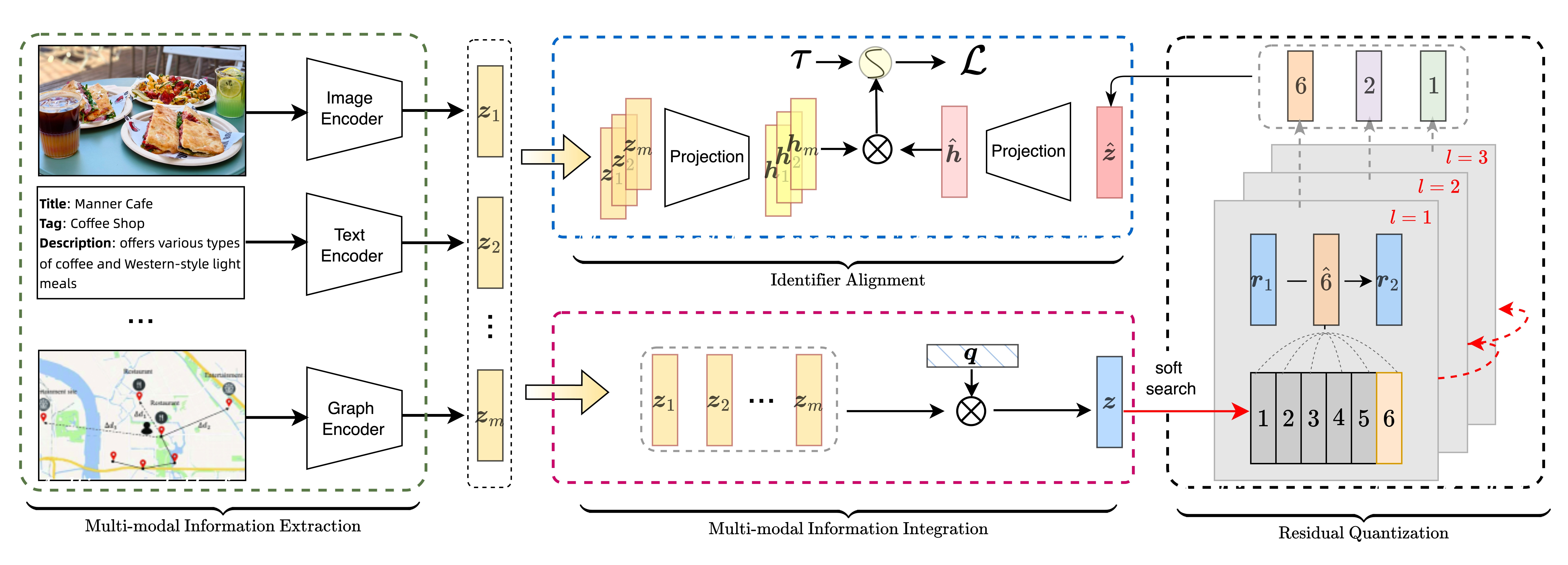}
    %\caption{\textbf{The framework of SimCIT. We set the fusion for an item as the training query $x_q$ and leave the other modal $x_k$ along with the views of other items as the keys in contrastive learning. Then, we extract the embeddings for these items and forward them to the quantization module to get quantized reconstructions. Embeddings reconstructed from the code memory bank serve as additional negative keys that boost contrastive learning. Next, we maximize the similarity between the query and the positive key and minimize the similarities of negative query-key pairs. Finally, we update the code memory bank with the quantization codes of the views of current image batch.}}
    \caption{\textbf{The framework of SimCIT. We employ an learnable attention module to integrate the multi-modal features. The Integrated representation is then fed to perform a soft residual quantization module. 
    Then the quantized reconstruction is taken as query to align with the all the modalities of the items, achieved by a set of contrastive loss.}}
    \label{fig:enter-label}
\end{figure*}

% These general MMCL frameworks provide a strong theoretical and practical foundation for multi-modal information fusion in POI contexts.
% In the recommendation domain, contrastive learning has been widely used to improve user and item representations, address data sparsity, and improve recommendation quality by modeling self-supervised signals from user interactions \cite{Wu2021SGL, Zhou2021CLRec}. 
% Closer to the approach taken in this paper, many works have begun exploring the application of contrastive learning to acquire more discriminative discrete representations, such as deep-hashing and quantization. In deep shingling, contrastive losses have been utilized effectively to learn binary codes for efficient similarity retrieval \cite{Zhang2019DeepHash, Cao2018DeepSupervisedHashing}. This approach highlights the importance of learning discrete codes with strong discriminative capabilities, which aligns well with the objective of generating semantic IDs to avoid collisions.

Recently, several works study contrastive learning from the perspective of information theory. Tian et al.\cite{10.5555/3495724.3496297} argues mutual information between views should be reduced by data augmentation while keeping task-relevant information intact. Wu et al.\cite{wu2020micl} shows the family of contrastive learning algorithms maximizes a lower bound on mutual information between multi-views where typical views come from augmentations, and finds the choice of negative samples and views are critical to these algorithms. We
build upon this observation with an optimization framework for selecting contrastive modalities of the items.

% Within this context, Contrastive Quantization with Code Memory (MeCoQ) \cite{Wang_Zeng_Chen_Dai_Xia_2022} introduced an unsupervised deep quantization solution that leverages contrastive learning instead of reconstruction losses to learn binary descriptors. This work emphasized the importance of regularization of codeword diversity to prevent model degeneration and highlighted the ability to capture discriminative visual semantics through contrastive learning. Building upon this, CoST \cite{Zhu2024} proposed a contrastive quantization-based semantic tokenization approach specifically for generative recommendation. CoST learns semantic tokens by harnessing both item relationships and semantic information, demonstrating significant improvements in generative recommendation performance and underscoring the critical impact of semantic tokenization on these models \cite{Zhu2024}.

\section{Proposed Method}
\label{sec:proposed_method}
In this section, we 
% introduce some preliminaries for our method, as a critical step for generative retrieval tasks.
% Then we outline the pipeline of our method, SimCIT, which consists of three primary components: extracting multi-modal information of the items from different domains, fusing the multi-modal semantic information into the identifiers, aligning the identifiers with the multi-modal semantics in an end-to-end way. A description of its integration with the generation task is finally followed.
% Each component will be discussed in detail in the subsequent subsections.
introduce our proposed approach for item tokenization, as a critical step for generative retrieval tasks.

\subsection{Problem Formulation}

% In a common sequential recommendation setting,
We consider the sequential recommendation task, where the item set $\mathcal{I}=\{i_n\}_{i=1}^{N}$. 
A user’s historical interactions are denoted as an item ID sequence $S = \left[i_1, i_2,..., i_n \right]$ arranged in chronological order, where $i \in \mathcal{I}$ denotes an interacted item. Sequential recommendation aims to capture user preferences and predict the next potential item $i_{n+1}$ based on $S$. Different from the conventional setting, in generative paradigm, each item is associated with a tuple of codes $(c_1,...,c_L)$ derived from its associated side features, like text and image, serving as its identifier with length $L$. 
Through the above process called item tokenization, the item sequence $S$ and the target item $i_{n+1}$ can be tokenized into a pair of code sequences, where each item is represented by $L$ codes. 
Thus, the sequential recommendation task is reformulated as a seq-to-seq problem, in which the next-item prediction is achieved by auto-regressively generating the target item identifier. Formally, this problem can be written as:
\begin{equation*}
    \label{eq:genrec}
    p(i_{n+1}|S) = \prod_{l=1}^L p(c_l^{n+1}|S,c_l^{n+1},...,c_{l-1}^{n+1}) 
\end{equation*}
Specifically, identifying the next token is typically approached as a discriminative problem, which require to capture the relationships among the items.  
However, to assign item identifiers, previous methods usually adopt a reconstruction-based structure, which aims to precisely reconstruct item embedding vectors using a MSE loss. In this paper, we address this limitation by learning a contrastive item
tokenizer. 

% Specifically, identifying the next token is typically approached as a classification problem, which necessitates maintaining a limited vocabulary while ensuring well-structured relationship of items. — dual requirements that fundamentally motivate our approach by enabling (1) discriminative identifier retrieval through sparse yet informative codeword distributions, and (2) collision-resistant semantic preservation via geometrically regularized embedding alignment

\subsection{Multi-modal Semantic Fusion}
To develop the universal item tokenizer, a key step is to sufficiently leverage the rich item semantics and establish a representation scheme that can well transfer across domains.
% Unlike previous text-based transferable item representations. we consider fusing multi-modal item content for capturing more comprehensive semantic information. 
% Moreover, to achieve multi-domain fusion item tokenization, we design a novel representation discretization
% approach via codebooks.
% In large-scale recommendation scenarios, multiple features are widely used to model the relationship among the massive items, 
% such as visual pictures, textual descriptions, geographical locations, and collaborative filtering signals. 
Throughout this work, we mainly consider the following modalities of the items: image,text,collaborative signals and spatial relationships, which critical in POI recommendation.
For text and image, we employ a frozen modal encoder(LLaMa or ViT \cite{dosovitskiy2020image}) followed by a MLP to encode item content, and obtain the item representations. 

For collaborative signals or spatial relationships, we employ graph encoders to extract the items' representations which corresponds to the node embeddings in graph.
Specifically, we describe a entity in terms of its items, by building two graphs: 
distance graph and check-in graph, where the edge weights are computed as distance and human check-in connectivity, respectively. We then apply graph encoder, such as \cite{10.5555/3294771.3294869}, as the representation learner to learn embeddings over these two constructed graphs respectively. 

Thus, we obtain a set of representations from different modalities: $\left\{\boldsymbol{z}_m\right\}_{m=1}^{\mathcal{M}}$, where $\mathcal{M}$ denotes the number of available modalities.
We argue that
representations derived from multiple modalities convey both shared and complementary information. However, existing methods for fusing representations tend to rely on simple techniques such as summation or concatenation. As a result, important informative may be ignored. To tackle this problem, we compute the importance of modality $m$ as follows:
\begin{equation*}
    % p_m = \frac{\exp\left(\boldsymbol{z}_m \cdot \boldsymbol{h} \right)}{\sum_{j=1}^{M}\exp\left(\boldsymbol{z}_j \cdot \boldsymbol{h} \right)}
    p_m = \text{softmax} \left( \boldsymbol{q}^T \boldsymbol{z}_m \right)
\end{equation*}
where $\boldsymbol{q}$ denotes the attention vector and $p_m$ represents the importance score for the representation of a single modality. Note that these parameters are shared for all modalities. Then, the modality-common embedding can be represented as
\begin{equation}
 \boldsymbol{z} = \sum_{m=1}^{|\mathcal{M}|} p_m \cdot \boldsymbol{z}_m
\end{equation}

To facilitate aligning the multi-modal semantics, we propose learn a codebook with residual quantization.
% With codebooks, we encode image, text and other modalities into a joint
% vision-language embedding space and learn the alignment by contrasting their prototype assignments. The codebooks can also be interpreted as underlying feature distribution for the paired data. In this way, by aligning features from each modality with the codebook, we implicitly align multi-modal features indirectly. In other words, the codebooks serves as a ``bridge" between the modalities.

\subsection{Contrastive Item Tokenization}
% In this section, we introduce an item tokenization method based on contrastive learning that integrates multi-source item features. Suppose we have M different data sources, \( x_m \). After undergoing feature extraction as described in the previous section, we obtain M multi-modal embeddings, \( z_m \), which represent multiple perspectives of an item. The integrated representation is obtained by a weighted fusion of these embeddings: 
% \begin{equation}
%  \boldsymbol{z} = \sum_{i=1}^{M} p_i \cdot z_i
% \end{equation}
% where \( p_i \) represents the weight for the representation of a single perspective, calculated as follows:
% \begin{equation*}
%     p_i = \frac{\exp\left(z_i \cdot h\right)}{\sum_{j=1}^{M}\exp\left(z_j \cdot h \right)}
% \end{equation*}
% In this section, we introduce a novel item tokenization method based on the contrastive learning framework to fully exploit the information embedded in multiple views. Suppose that we have $M$ views of the items, denoted by $\mathrm{x}_m$. 
% Then the fused representation $z$ can be as
% \begin{equation}
%     z = \sum_{m=1}^{M}p_m z_m
% \end{equation}
% Next, we perform residual quantization on the obtained fused representation \( \boldsymbol{z} \). 
Residual quantization have been validated in various related works, typically accompanied by a reconstruction objective and a commitment loss concerning the codebook constraint. In this section, we introduce a novel quantization representation learning framework that drops the traditional VAE-based architecture and its associated losses in favor of a contrastive learning approach.

% Residual quantization is an iterative process that quantizes the residual vector at each step, thereby generating a tuple of tokens. 
We firstly establish a set of learnable codebooks $\boldsymbol{C}_l = \left\{\boldsymbol{e}_k^l|k=1,...,K\right\}, l \in [1,L]$, where $L$ represents the number of codebooks and $K$ denotes the size of each codebook. 
Residual quantization is an iterative process that quantizes the residual vector at each step, thereby generating a tuple of tokens. 
Initially, the residual is defined as $\boldsymbol{r}_0 = \boldsymbol{z}$. 
In the usual setting, at each level $l$, to approximate $\boldsymbol{r}_l$, 
the residual quantization process can be represented as:
\begin{equation}
\begin{gathered}
    \label{eq:hard_assign}
    c_l = \arg\min_k \left\| \boldsymbol{r}_l - \boldsymbol{e}_k^l \right\|_2^2, \\
    \boldsymbol{r}_{l+1} = \boldsymbol{r}_l - \boldsymbol{e}_{c_l}^l,
\end{gathered}
\end{equation}
% a nearest neighbor search is performed to
% %using the first codebook $C_1$, 
% select the closest code vector with its index $c_l = \arg \min_k \|r_l - \mathbf{e}_l^k\|_2$ and the residual at the next level is obtained by $r_{l+1} = r_l - \mathbf{e}^l_{c_l}$.
By iteratively repeating this procedure $L$ times, we generate a tuple of $L$ codes $(c_1,...,c_L)$, which are denoted as semantic tokens.
However, the nearest neighbor search operation $\arg\min$ is non-differentiable and can't be optimized by performing backpropagation.

To enable gradient propagation through the non-differentiable nearest neighbor search operation $\arg\min$, we adopt a categorical reparameterization with Gumbel-softmax \cite{jang2017categoricalreparameterizationgumbelsoftmax,9729603} to relax the discrete optimization of hard codeword assignment to a trainable form. Specifically, in the training process,  we approximate $\arg\min$ via softmax with a temperature $\alpha$ at level $l$, 
% the residual quantization can be represented as,
% is a technique that allows sampling from categorical distribution during the forward pass of a neural network. It essentially is done by combining the reparameterization trick and smooth relaxation. Let’s look at how this works.
% gumbel-softmax reparam
% the residual quantization process can be represented as:
\begin{equation}
\begin{gathered}
    d_k^l= -\left\|\boldsymbol{r}_l-\boldsymbol{e}_k^l\right\|_2^2 + \epsilon_l \\
    c_l^k = \frac{\exp(d_k^l/\alpha)}{\sum_{k=1}^K \exp(d^l_k/\alpha)}
    % \text{softmax}_{k}(\boldsymbol{g}_k^l/\alpha)
\end{gathered}
\end{equation}
% \begin{equation}
%     \boldsymbol{a}_l = \text{softmax}(g_l/\alpha)
%     % \frac{\exp(g_l/\alpha)}{\sum_{j=1}^K \exp(g_l/\alpha)}
% \end{equation}
where $c_l^k$ is the assignment weight of the residual vector $\boldsymbol{r}_l$ relative to the codeword $\boldsymbol{e}_k^l$. Gumbel-distributed noise $\epsilon_l \sim \text{Gumbel}(0,1)$ follows an annealing schedule in the training process: initialized at $\alpha > 0$ for exploration of the codebook $\boldsymbol{C}_l$, then exponentially decayed to approximate 0 (hard assignment). During forward propagation, we get the residual at the next level through the following steps,
\begin{equation}
\label{eq:soft_assign_2}
    \boldsymbol{r}_{l+1} = \boldsymbol{r}_l - \sum_{k=1}^K c_l^k \cdot \boldsymbol{e}_k^l
\end{equation}
% Finally, we get soft quantization code and the soft quantized reconstruction of $z$ by summing over $L$ code vectors, 
% \begin{equation}
%     \hat{z} = \sum_{l=1}^{L}\mathbf{e}_{c_i}^{l}
% \end{equation}
% The softmax is a differentiable alternative to the nearest neighbor search $\arg\max$ that relaxes the discrete optimization of hard codeword assignment to a trainable form.
% To approximate $r_0$, the residual quantization process can be represented as:
% we perform a nearest neighbor search using the first codebook $C_1$, selecting the closest code vector with its index $c_1 = \arg \min_k \|r_0 - e_1^k\|_2$. Next, we compute the residual at the second level, $r_1 = r_0 - e_1^{c_1}$, which captures the information not represented by the first code vector. This residual is then approximated using the second codebook $C_2$. By iteratively repeating this procedure $L$ times, we generate a tuple of $L$ codes $(c_1,...,c_L)$, which are denoted as semantic tokens.
% Specifically, given the semantic tokens after quantization, we construct an approximated vector $\hat{z}$ by summing over $L$ code vectors: $\hat{z} = \sum_{i=1}^{L}e_i^{c_i}$.

By iteratively repeating this procedure $L$ times, we get the reconstructed embedding as,
\begin{equation}
\label{eq:soft_assign_3}
    \hat{\boldsymbol{z}} = \sum_{l=1}^{L}\sum_{k=1}^{K} c_l^k \cdot \boldsymbol{e}_k^l
\end{equation}
compared to Eq.(\ref{eq:hard_assign}), our differentiable learning strategy enables full gradient backpropagation to entire codebooks during training with learnable soft assignment weights, thereby circumventing the gradient blocking issue inherent in conventional hard vector quantization approaches. 

\begin{algorithm}
    \caption{SimCIT's main learning algorithm.}
    \label{alg:simclr}
    \begin{algorithmic}[1]
        \STATE \textbf{input:} constant $\tau$, $\alpha$, structure of $f_m$, $g$, codebook $\boldsymbol{C}$.
        \FOR{all modality input $\{\boldsymbol{x}_m\}_{m=1}^\mathcal{M}$}
            \STATE \textcolor{gray}{\# extract multi-modal representations}
            \STATE $\boldsymbol{z}_m = f_m(\boldsymbol{x}_m)$ \textcolor{gray}{\# representation}
            \STATE $\boldsymbol{h}_m = g(\boldsymbol{z}_m)$ \textcolor{gray}{\# projection}
        \ENDFOR
        \STATE \textcolor{gray}{\# fuse the representations}
        \vspace{3pt}
        \STATE $\boldsymbol{z} = \sum_{m=1}^{|\mathcal{M}|} p_m \cdot \boldsymbol{z}_m$
        \vspace{3pt}
        \STATE \textcolor{gray}{\# apply soft residual quantization}
        \FOR{all $l \in \{1,...,L\}$}
        \STATE $d_k^l= -\left\|\boldsymbol{r}_l-\boldsymbol{e}_k^l\right\|_2^2 + \epsilon_l$ \textcolor{gray}{\# distance}
        \vspace{4pt}
        \STATE $c_l^k = \text{softmax}_k(d_k^l/\alpha)$  \textcolor{gray}{\# assignment weight}
        \vspace{4pt}
        \STATE $\boldsymbol{r}_{l+1} = \boldsymbol{r}_l - \sum_{k=1}^K c_l^k \cdot \boldsymbol{e}_k^l$ \textcolor{gray}{\# residual vector}
        \ENDFOR
        \STATE $\hat{\boldsymbol{z}} = \sum_{l=1}^{L}\sum_{k=1}^{K} c_l^k \cdot \boldsymbol{e}_k^l$ \textcolor{gray}{\# reconstructed representation}
        \vspace{4pt}
        \STATE $\hat{\boldsymbol{h}} = g(\hat{\boldsymbol{z}})$ \textcolor{gray}{\# projection} 
        \STATE define loss: $\mathcal{L} = - \sum_{m=1}^{|\mathcal{M}|}\log \frac{\exp \left(\hat{\boldsymbol{h}} \cdot \boldsymbol{h}_m^+ /\tau \right) }{\sum_{\boldsymbol{h}^-\in \mathcal{B}} \exp\left(\hat{\boldsymbol{h}}\cdot \boldsymbol{h}^{-} /\tau \right)}$
        \vspace{4pt}
        \STATE update networks $f_m$, $g$ and codebooks $\boldsymbol{C}$ to minimize $\mathcal{L}$
        \STATE \textbf{return} encoder network $f_m(\cdot)$ and codebooks $\boldsymbol{C}$, and throw away $g(\cdot)$
    \end{algorithmic}
\end{algorithm}

For learning a more discriminative item identifier, we introduce a contrastive learning-based method, which is significantly different from the popular reconstruction-based VAE technique.
Specifically, we enforce the \textit{NT-Xent} loss \cite{Chen2020SimCLR} on the reconstructed embedding with all the multi-modal representations. 
Similar to \cite{Chen2020SimCLR}, we instead compute the loss on the representations after passing through a projection head $g$, denoted as $\boldsymbol{h} = g(\boldsymbol{z})$. The final objective is as follows, 
\begin{equation}
\label{eq:loss}
    % \mathcal{L} = - \sum_{m=1}^{|\mathcal{M}|}\log \frac{\exp \left( \hat{\boldsymbol{z}} \cdot \boldsymbol{z}_m^+ /\tau \right) }{\exp\left(\hat{\boldsymbol{z}}\cdot \boldsymbol{z}^{+}_{m}/\tau\right) +\sum_{\boldsymbol{z}^-\in \mathcal{B}} \exp\left(\hat{\boldsymbol{z}}\cdot \boldsymbol{z}^{-} /\tau \right)}
    \mathcal{L} = - \sum_{m=1}^{|\mathcal{M}|}\log \frac{\exp \left(\hat{\boldsymbol{h}} \cdot \boldsymbol{h}_m^+ /\tau \right) }{\sum_{\boldsymbol{h}^-\in \mathcal{B}} \exp\left(\hat{\boldsymbol{h}}\cdot \boldsymbol{h}^{-} /\tau \right)}
\end{equation}
where $\tau$ denotes a temperature hyper-parameter and $\mathcal{B}$ is item representations for negative items within a batch. 
We claim that this is the final objective function without any other regularization terms. 
By optimizing the proposed loss function, it enables end-to-end training of both the model itself and the entire codebooks. This dual learning mechanism empowers item identifiers to learn multiple modalities shared information and enhance discriminative capacity by pushing different items apart.

With the above process, we encode image, text and other modalities into a joint
vision-language embedding space and learn the alignment by contrasting their prototype assignments. The codebooks can also be interpreted as underlying feature distribution for the paired data. In this way, by aligning features from each modality with the codebooks, we implicitly align multi-modal features indirectly. In other words, the codebooks serves as a ``bridge" between the modalities (Fig. \ref{fig:background}).

% Since the
% batch data is constructed randomly, the negative instances within a batch are from a mixture of multiple domains, thereby facilitating the fusion of cross-domain knowledge.

\subsubsection{Implicit Regularization to Promote Diversity}
% \subsubsection{}
In previous works \cite{10.1145/3627673.3679569} \cite{10.1145/3640457.3688178}, a penalty function is usually utilized to promote the diversity of the identifiers. 
In this part, we demonstrate that the proposed loss function $\mathcal{L}$ implicitly promote the diversity of the identifiers. Specifically, the batch of representations for negative item $\mathcal{B}$ can be divided into 2 parts: the other items within the reconstructed representations $\mathcal{B}_c$ and the counterpart of certain modality $\mathcal{B}_m$.
Therefore, the equation Eq.\ref{eq:loss} can be reformulated as:
\begin{equation*}
    \mathcal{L} = - \sum_{m=1}^{|\mathcal{M}|} \log \frac{\exp \left( \hat{\boldsymbol{h}} \cdot \boldsymbol{h}_m^+ /\tau \right) }{\sum_{\hat{\boldsymbol{h}}^-\in \mathcal{B}_c} \exp\left(\hat{\boldsymbol{h}}\cdot \hat{\boldsymbol{h}}^{-} /\tau \right) +C_{\mathcal{B}_m}}
\end{equation*}
where $C_{\mathcal{B}_m}$ in denominator represents the component about modality alignment. 
Therefore, by ignoring other negative samples in $\mathcal{B}_m$, this formulation can be considered as bringing the identifier closer to the anchor point $\boldsymbol{z}_m$ while maximizing the dispersion among identifiers of different items. The empirical results in section \ref{sec:experiments} demonstrate this perspective.

% \subsubsection{Quantization Queue}
% Effective contrastive learning methods rely on sufficient negative samples to learn discriminative representations. Therefore, existing
% memory-based methods cache the image embeddings and serve them as the negatives in the later training. However, as the model keeps updating, the early cached embeddings expire and become noises to CL. To enhance the effect of memory, MoCo introduces a momentum encoder that mitigates the embedding aging issue at a higher computation cost. In contrast, we find an elegant solution that achieves a similar effect more efficiently.

\subsubsection{Minimal Sufficient Identifier}
\begin{figure}
    \centering
    \includegraphics[width=\linewidth]{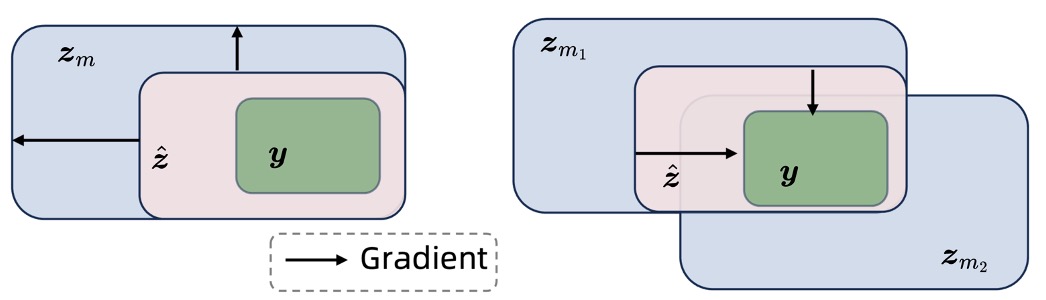}
    \caption{Schematic of learning process of (left) VAE-based quantization representation learning and (right) multi-view contrastive representation learning.}
    \label{fig:mini_suff}
\end{figure}

% reconstruction loss and contrastive loss difference 
In this section, we investigate the difference of the representation learned by the contrastive learning from the traditional VAE-based architectures.
For simplicity, we take the modalities of the item as different views, which convey both shared and complementary information. 
In the context of generative retrieval tasks, the goal is to identify the unique item from the vocabulary, in which the take label is denoted as $\boldsymbol{y}$.
We suppose that we can recognize $\boldsymbol{y}$ according to the view $\boldsymbol{z}_m$, which is usually described as that the views $\boldsymbol{z}_m$ are sufficient for $\boldsymbol{y}$ \cite{10.5555/3495724.3496297, wu2020micl}. 
That is,
\begin{equation*}
    I(\boldsymbol{z}_{m_i};\boldsymbol{z}_{m_j}) \geq I(\boldsymbol{z}_{m_i};\boldsymbol{y}), \forall m_i,m_j \in \{1,...,\mathcal{M}\}
\end{equation*}
where $I(\cdot;\cdot)$ denotes mutual information. Therefore, our proposed loss function can be taken as finding the minimal identifier to identify $y$, that is,
\begin{equation*}
    \mathcal{L} = \min_{\hat{\boldsymbol{z}}}I(\left\{\boldsymbol{z}_m\right\}_{m=1}^{\mathcal{M}}; \hat{\boldsymbol{z}})
\end{equation*}
by introducing more sufficient modalities of the items, $\hat{\boldsymbol{z}}$ gradually converge to a minimal sufficient identifier. The Schematic of learning process for SimCIT are illustrated in Fig. \ref{fig:mini_suff}.

\subsection{Autoregressive Generation}
Following the TIGER model, we employ an encoder-decoder transformer architecture for generative recommendation. Specifically, we tokenize each item into a tuple of tokens, resulting in a behavior sequence $S = (c_{1}^1,..., c_{L}^1, c_{1}^2,..., c_{L}^2,..., c_{L}^N)$, where $N$ represents the sequence length of items. The model is then trained as a seq2seq learning task for autoregressive token generation. Given the token sequence of the first $i - 1$ items, the model predicts the tokens $(c_{1}^i,..., c_{L}^i)$ for the $i-th$ item $t_i$ autoregressively, rather than generating an item ID directly. After training, we use beam search to generate tokens and then lookup the token-item table to obtain the corresponding items for recommendation.

\section{Experiments}
\label{sec:experiments}
We conduct a comprehensive set of experiments to validate the effectiveness of our method in various settings, addressing several key research questions:
\begin{itemize}
    \item \textbf{RQ1}: How does SimCIT enhance performance compared to baseline methods in standard sequential recommendation tasks?
    \item \textbf{RQ2}: What impact does our diverse token generation strategy and collision avoidance strategy have?
    \item \textbf{RQ3}: How does varying the training intensity of item-identifier alignment influence item identifier refinement, and what are the effects on overall sequential recommendation task performance?
    \item \textbf{RQ4}: In what ways does SimCIT enhance item identifier quality?
\end{itemize}

Specifically, the experiments are organized as follows: in section \ref{experiments_setup}, we firstly introduce the experiments setup, mainly including the utilized datasets and baseline methods.
In section \ref{sec:overall_performance}, we provide a detailed comparison with other baseline methods on the datasets. Section \ref{sec:ablation} shows the ablation study of our proposed method and further analysis are illustrated in the last section \ref{further_analysis}.

\subsection{Experiments Setup}
\label{experiments_setup}
\subsubsection{Dataset}
We utilize four real-world public recommendation datasets and one large scale industrial dataset from different domains: 
\begin{itemize}
    \item \textbf{Instruments} \cite{ni-etal-2019-justifying} (abbr. INS) is from the Amazon review datasets, which contains user interactions with rich music gears.
    \item \textbf{Beauty} \cite{ni-etal-2019-justifying} (abbr. BEA) contains user interactions with extensive beauty products from Amazon review datasets.
    \item \textbf{Foursquare} \cite{6844862} is collected over 11 months, comprise users' historical POI trajectories from New York City (denoted as NYC) and Tokyo (denoted as TKY), sourced from Foursquare.
\end{itemize}

However, we argue that the user interaction behavior length in these public datasets are too short and the number of item tokens is insufficient to demonstrate the effectiveness of generative recommendation. Therefore, we conduct experiments on a large-scale industrial dataset.
\begin{itemize}
    \item \textbf{AMap}$^{1}$: collected over 6 months, recording the users' historical poi trajectories from Beijing, China. 
\end{itemize}

\footnote{$^1$ https://amap.com/}
% 1) \textbf{Instruments} is from the Amazon review datasets [29], which contains user interactions with rich music
% gears. 2) Beauty contains user interactions with extensive beauty
% products from Amazon review datasets [29].
% : Foursquare-NYC, Foursquare-TKY \cite{}, and XXX [5]. The first two datasets, collected over 11 months, comprise data from New York City and Tokyo, sourced from Foursquare.
% We utilize data that has been preprocessed as per the methods detailed by \cite{}. 
Following previous studies, all the datasets are preprocessed as follows: (1). Filter out items with fewer than 10 visit records in history; (2). Exclude users with fewer than 10 visit records in history; 
% (3). Divide user check-in records into several trajectories with 24-hour intervals, excluding trajectories that contain only one check-in.
The check-in records are sorted chronologically: the first 80\% are used for the training set, the middle 10\% are defined as the validation set, and the last 10\% are defined as the test set. Note that the validation and test set has to contain all users and items that appear in the training set. The detailed statistics of preprocessed datasets are presented in Table \ref{label:stat}.
\begin{table}[h!]
\centering
\setlength{\tabcolsep}{8pt} % Adjusts column separation
\renewcommand{\arraystretch}{1.2} % Adjusts row separation
\caption{Statistics of the preprocessed datasets. Avg.len denotes the average length of item sequences.}
\label{label:stat}
\begin{tabular}{lccccc}
\toprule[1pt]
\textbf{Dataset} & \textbf{\#Users} & \textbf{\#Items} & \textbf{\#Inter.} & \textbf{Avg.len} & \textbf{Sparsity} \\
\midrule
% Pre-training & 999,334 & 344,412 & 8,609,909 & 8.62 & 99.997\% \\
INS & 57,439 & 24,587 & 511,836 & 8.91 & 99.9640\% \\
BEA & 50,985 & 25,848 & 412,947 & 8.10 & 99.9690\% \\
NYC & 1,075 & 5,099 & 104,074 & 96.8 & 98.1013\%   \\
TKY & 2,281 & 7,844 & 361,430 & 158.4 & 97.9798\%  \\
AMap & 7,684k & 6,158k & 172,100k & 22.39 & 99.9996\%  \\
\bottomrule[1pt]
\end{tabular}
\end{table}

\subsubsection{Baselines Methods}
To enable a comprehensive comparison, we categorize the baseline methods into 2 distinct groups.
\begin{itemize}
    \item \textbf{Traditional Sequential Recommendation Models}: In E-commerce recommendation task, GRU4Rec \cite{hidasi2016sessionbasedrecommendationsrecurrentneural} is originally designed for session-based recommendation, using GRUs to model the sequence of user interactions. SASRec \cite{8594844} leverages self-attention mechanisms to capture long-term dependencies in user interaction histories. BERT4Rec \cite{10.1145/3357384.3357895} adopts a Bert-style bidirectional language model to learn item representations. For poi recommendation task, we compare our model with the following baselines: STGCN \cite{9133505} incorporates gating mechanisms to effectively model temporal and spatial intervals in check-in sequences, thereby capturing both short-term and long-term user preferences. STAN \cite{10.1145/3442381.3449998} leverage a bi-layer attention architecture and aggregates spatio-temporal correlations within user trajectories, learning patterns across both adjacent and non-adjacent locations as well as continuous and non-continuous visits. GeoSAN \cite{10.1145/3539618.3591770} constructs a hypergraph transformer to capture inter and intra-user relations, which solves the cold-start problem.
    \item \textbf{LLM-Based Recommendation with Item Tokenization}: These methods utilize more sophisticated item tokenization methods. TIGER \cite{Rajput2023} employs a codebook-based approach using RQ-VAE, quantizing item embeddings into code sequences. LETTER \cite{10.1145/3627673.3679569} similarly utilizes RQ-VAE for generating item identifiers, incorporating collaborative and diversity regularization during training. These approaches initialize items with new token sequences, also serving as item identifier initialization method within our method.
\end{itemize}

% \textbf{Sequential Recommenders}:
% \begin{itemize}
%     \item GRU4Rec \cite{} leverages Gated Recurrent Units (GRUs) to
% model sequential patterns in user interactions.
%     \item BERT4Rec \cite{} employs the bidirectional self-attention mech-
% anism with a masked prediction objective for sequence modeling.
%     \item SASRec \cite{} utilizes a unidirectional self-attention network
% to model user behaviors.
%     \item FMLP-Rec \cite{} introduces an all-MLP model with learnable
% filters to reduce noise and effectively capture user preferences.
% \end{itemize}
% \textbf{Generative Recommenders}:
% \begin{itemize}
%     \item TIGER \cite{} employs RQ-VAE to map items into semantic IDs,
% which serve as item identifiers, and adopts the generative retrieval
% paradigm for sequential recommendation.
%     \item LETTER \cite{} enhances TIGER by incorporating collaborative
% and diversity regularization into RQ-VAE.
%     \item LLM4POI \cite{} ...
% \end{itemize}
\subsubsection{Evaluating Settings}
We evaluate model performance using a widely adopted metrics: top-$K$ recall, with $K$ set to (5, 10).
% Following previous studies, we employ the leave-one-out strategy for dataset splitting. For each user interaction sequence, the latest item is designated as the test data, the second most recent item as the validation data, and all remaining items as the training data. 
To ensure a rigorous comparison, we conduct the full-ranking evaluation over the entire item set. Additionally, the beam size is set to 50 for all generative recommendation models.

\begin{table}[h]
\centering
\setlength{\tabcolsep}{1.5 pt} % Adjusts space between columns
\renewcommand{\arraystretch}{1.2}
\caption{The overall performance comparisons between different baseline methods and SimCIT. The best and second-best results are highlighted in bold and underlined font, respectively.}
\label{label:overall}
\begin{tabular}{llccc|ccc}
\toprule[1pt]
\multirow{2}{*}{Dataset} & \multirow{2}{*}{Metric} & 
\multicolumn{3}{c}{Sequential Recommendation} & \multicolumn{3}{c}{Generative Recommendation} \\
\cmidrule(lr){3-5}
\cmidrule(lr){6-8}
& & \scriptsize{GRU4Rec} & \scriptsize{BERT4Rec} & \scriptsize{SASRec} & \scriptsize{TIGER} & \scriptsize{LETTER} & \scriptsize{SimCIT} \\
\midrule
\multirow{2}{*}{INS} & Rec@5 & 0.0624 & 0.0671 & 0.0751& 0.0870 & \underline{0.0913} &  \textbf{0.0950} \\
& Rec@10 & 0.0701 & 0.0822 & 0.0947 & 0.1058 & \underline{0.1122} & \textbf{0.1195} \\
\midrule
\multirow{2}{*}{BEA} & Rec@5 & 0.0202 & 0.0203 & 0.0380 & 0.0395 & \underline{0.0431}  & \textbf{0.0466} \\
& Rec@10 & 0.0338 & 0.0347 & 0.0588 & 0.0610 & \underline{0.0672}  & \textbf{0.0693} \\
\midrule[0.5pt]
% \multirow{2}{*}{} & \multirow{2}{*}{} & 
% \multicolumn{3}{c}{Sequential Recommendation} & \multicolumn{3}{c}{Generative Recommendation} \\
% \cmidrule(lr){3-5}
% \cmidrule(lr){6-8}
& & \scriptsize{STGN} & \scriptsize{GeoSAN} & \scriptsize{STAN} & \scriptsize{TIGER} & \scriptsize{LETTER} & \scriptsize{SimCIT} \\
\midrule
\multirow{2}{*}{NYC} & Rec@5 & 0.2439 & 0.4006 & 0.4669 & 0.4812 & \underline{0.4851} & \textbf{0.5021} \\
& Rec@10 & 0.3015 & 0.5267 & 0.5962 & 0.6122 & \underline{0.6281} & \textbf{0.6473} \\
\midrule
\multirow{2}{*}{TKY} & Rec@5 & 0.1940 & 0.2957 & 0.3461 & 0.3688 & \underline{0.3961} & \textbf{0.4075} \\
& Rec@10 & 0.2710 & 0.3740 & 0.4264 & 0.4617 & \underline{0.4701} & \textbf{0.4852} \\
\bottomrule[1pt]
\end{tabular}
\end{table}

\subsubsection{Implementation Details}
In the experiments on E-commerce dataset (BEA and INS), we integrated two sources of embedding data:

a. \textit{Textual Descriptions}: We utilized the pretrained BERT model to merge the textual information of items into an embedding, which encapsulates the semantic information of the items. 
% For instance, for the case of an Arts \& Crafts Store, the most related entities are Gift Shop, Miscellaneous Shop, and more.

b. \textit{Collaborative Signals}: We employed the alternating least squares \cite{10.1145/2365952.2365972, 10.1145/2911451.2911489} method to obtain the item embeddings. We established a matrix representation of the relationships between users and items, where the number of interactions between users and items is treated as rating data, resulting in a 32-dimensional item embedding. This data represents the collaborative filtering relationship embedding.

In POI recommendation tasks (TKY, NYC and AMap), We've leveraged the text and collaborative filtering signals mentioned above. Given the unique characteristics of POI (Point of Interest) recommendation, we also incorporated the following image features (available in AMap) and geographical graph features. 

c. \textit{Visual Descriptions}: We use the pretrained ViT model to convert an item's image information into an embedding, which serves as the item's image feature.

d. \textit{Spatial Graph}: We measured spatial relationship intensity using the distance between POI points, constructing a spatial relationship graph between POIs. Subsequently, we applied GraphSAGE \cite{10.5555/3294771.3294869} to obtain the spatial relationship embedding the embedding.

With regard to the hyperparameter settings, the SimCIT model consists of three
core components: a MLP encoder, a residual quantizer, and a projection layer. The encoder includes three hidden layers with dimensions [512, 256, 128], utilizing ReLU activations, and outputs a 96-dimensional latent representation. Each item is assigned a unique 3-tuple semantic token, with shared codebooks across the three levels, and the codebook size is set to 48 (128 for AMap because of its sheer scale of items.). The hyperparameters are set as $\alpha$ = 0.1 and $\tau$ = 0.1. We utilize the Adam optimizer with a learning rate of 0.0001 and a batch size of 256 for codebook learning. For seq2seq transformer model training, we adopt the same structure as TIGER, setting the batch size to 512 and the learning rate to 0.001.

\subsection{Overall Performance}\label{sec:overall_performance}
In this section,
we firstly compare SimCIT with various baseline models on public E-commerce recommendation benchmarks (INS and BEA dataset from Amazon). 
% including E-commerce (INS and BEA dataset from Amazon) and POI recommendation (NYC and TKY dataset from Foursquare). 
The overall results are shown in Table \ref{label:overall}. From these results, we have the following observations: 
Compared to traditional recommendation models (i.e., GRU4Rec, BERT4Rec, SASRec), 
Generative recommenders (i.e., TIGER, LETTER and SimCIT) benefit from the generative paradigm and the prior semantics within item identifiers, outperforming traditional recommendation models. Among these, SimCIT surpasses TIGER due to its application of collaborative signals in the item tokenizer. 
% Furthermore, limited by the domain-specific tokenizers and recommenders, generative baseline models do not consistently perform better than transferable sequential recommenders (e.g., MISSRec). 
% This further highlights the importance of developing a transferable generative recommender.

Furthermore, our proposed SimCIT maintains the best performance in all cases, exhibiting substantial improvements over traditional and generative baseline models. 
Different from previous generative recommenders, we propose a universal item tokenizer that leverages multi-modal content for item semantic modeling. Based on this, we develop a transferable generative recommender, which effectively enhances model performance by integrating the strengths of the generative paradigm and cross-domain knowledge transfer.

Secondly, for the results of POI recommendation task (NYC and TKY dataset from Foursquare), SimCIT achieves a larger improvement compared to the baseline methods than the E-commerce recommendation tasks. This demonstrates the critical role of spatial graph relationships for POI recommendation—a key strength of our proposed SimCIT framework—which maintains extensibility for arbitrary auxiliary information integration while preserving semantic topology awareness.

We ultimately validate our framework on a large-scale industrial dataset (TABLE \ref{tab:amap_results}). Having established its superiority over conventional sequential recommendation methods on public benchmarks, we now focus exclusively on LLM-based baselines. By extending the evaluation spectrum to larger $K$ thresholds ($K \in (10, 100, 1000)$), our method demonstrates consistent dominance across all Recall@$K$ metrics, achieving a 15\% improvement over the second-best performer—a critical advancement for real-world deployment where top-$K$ precision scalability is paramount. 
% \begin{table}[h!]
% \centering
% \setlength{\tabcolsep}{8pt} % Adjusts column separation
% \renewcommand{\arraystretch}{1.2} % Adjusts row separation
% \caption{Performance comparison in terms of Acc@1 on three datasets.}
% \begin{tabular}{l|c|c|c}
% \hline
% \textbf{Model} & \textbf{NYC} & \textbf{TKY} & \textbf{AMAP} \\
% \hline
% STGCN & 0.1799 & 0.1716 & / \\
% STAN & 0.2231 & 0.1963 & / \\
% STHGCN & 0.2734 & 0.2950 & / \\
% LLM4POI & 0.3372 & 0.3035 & / \\
% TIGER & / & / & /  0.1801\\
% LETTER & / & / & / \\
% \hline
% SimCIT & / & / & / \\
% \hline
% \end{tabular}
% \end{table}

\begin{table}[htbp]
\centering
\caption{Performance Comparison On Recall@$K$ Across Generative Recommendation Models on AMap dataset.}
\label{tab:amap_results}
\footnotesize
\setlength{\tabcolsep}{8pt}
% \begin{tabular}{@{}l rrrrr rrrr@{}}
\begin{tabular}{lcccc}
\toprule
\multirow{2}{*}{Methods}  & \multicolumn{3}{c}{AMap} \\
\cmidrule(lr){2-4}
% \toprule
%  Variants       & \multicolumn{4}{c}{AMap} \\
% \cmidrule(lr){2-5}
               & Recall@10 & Recall@100 & Recall@1000 \\
\midrule
TIGER.  & 0.2684 & 0.4510  & 0.7010 \\
LETTER   & 0.2758 & 0.4801  & 0.7210  \\
SimCIT   & \textbf{0.3206} & \textbf{0.5010} & \textbf{0.7827} \\
\bottomrule
\end{tabular}
% \vspace{2mm}
\end{table}

% \begin{table}[htbp]
% \centering
% \caption{Performance Comparison of ACC@1 on Different Datasets and Models}
% \label{tab:results}
% \small
% \setlength{\tabcolsep}{4pt}
% \begin{tabular}{@{}l *{6}{C} ccc@{}}
% \toprule
% \multirow{2}{*}{Dataset} & \multicolumn{6}{c}{Sequence Models} & \multicolumn{3}{c}{Spatial-Temporal Models} \\
% \cmidrule(lr){2-7} \cmidrule(l){8-10}
%  & GRU4Rec & BERT4Rec & SASRec & TIGER & LETTER & \textbf{SimCIT} & STGCN & STAN & STHGCN & LLM4POI & TIGER & LETTER & \textbf{SimCIT} \\
% \midrule
% Instruments & 0.32 & 0.35 & 0.38 & 0.42 & 0.45 & \textbf{0.51} & -- & -- & -- & -- & -- & -- & -- \\
% Beauty & 0.28 & 0.31 & 0.34 & 0.39 & 0.43 & \textbf{0.49} & 0.27 & 0.29 & 0.33 & 0.36 & 0.40 & 0.44 & \textbf{0.48} \\
% NYC & -- & -- & -- & -- & -- & -- & 0.25 & 0.28 & 0.31 & 0.35 & 0.38 & 0.42 & \textbf{0.47} \\
% TKY & -- & -- & -- & -- & -- & -- & 0.24 & 0.26 & 0.30 & 0.33 & 0.37 & 0.41 & \textbf{0.45} \\
% AMap & -- & -- & -- & \multicolumn{2}{c}{--} & \textbf{0.52} & -- & -- & -- & -- & 0.39 & 0.46 & \textbf{0.52} \\
% \bottomrule
% \end{tabular}
% \vspace{2mm}
% \footnotesize
% \textsuperscript{1} Bold numbers indicate our SimCIT results. "--" means not applicable. \\
% Sequence models include GRU4Rec, BERT4Rec, SASRec. Spatial-Temporal models include STGCN, STAN, STHGCN, LLM4POI. \\
% TIGER and LETTER are shared baseline models across categories.
% \end{table}
\begin{table}[ht]
\centering
\caption{Ablation study of SimCIT with different variants. Best results are highlighted in bold.}
\label{tab:ablation}
\begin{tabular}{lcccc}
\toprule
\multirow{2}{*}{Variants}  & \multicolumn{3}{c}{AMap} \\
\cmidrule(lr){2-4}
               & Recall@10 & Recall@100 & Recall@1000\\
\midrule
SimCIT w/o p.h.  & 0.2782 & 0.4313 & 0.7213 \\
SimCIT w/o g.s.  & 0.2253 & 0.3591 & 0.6097 \\
SimCIT w/o a.s.    & 0.2821 & 0.4716 & 0.7510 \\
SimCIT w/o m.f.  & 0.2809 & 0.4522 & 0.7022 \\
SimCIT  & \textbf{0.3206} & \textbf{0.5010} & \textbf{0.7827}\\
\bottomrule
\end{tabular}
\end{table}

\subsection{Ablation Study}
\label{sec:ablation}
% To investigate how the proposed techniques impact model performance, we conduct an 
% ablation study on AMap datasets. 
To investigate how the proposed techniques impact model performance, we conduct an ablation study on AMap dataset.

\begin{figure*}[h]
    \centering
    \includegraphics[width=\linewidth]{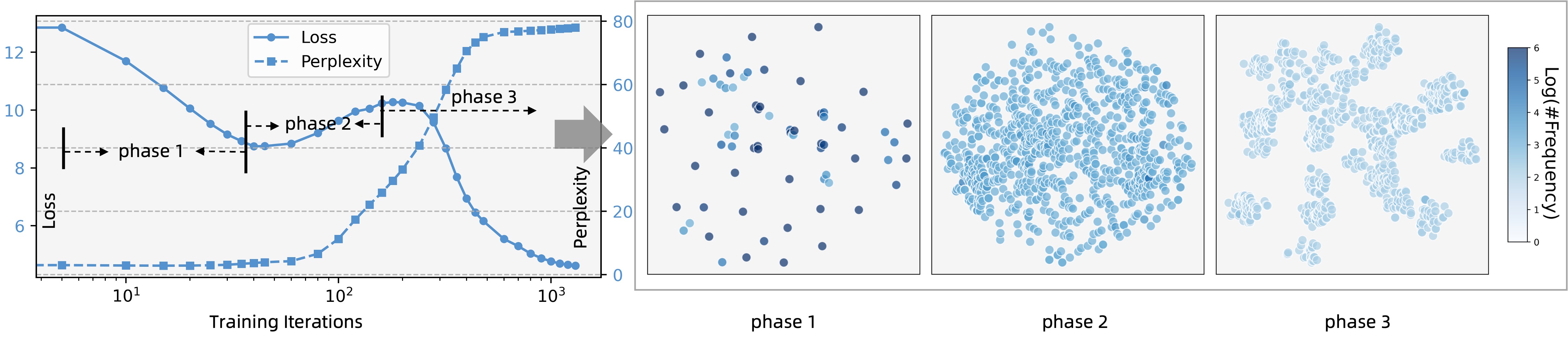}
    \caption{Training dynamics of the loss and perplexity of the identifier on the AMap dataset (left) and code embedding distributions of 3 phases (right). Each point represents a code embedding vector, and darker colors indicate that the code is assigned to more items. Three phases are observed in this process, phase 1: decreasing loss and high collision; phase 2: increasing loss associated with exploration of the codebooks and increase of diversity; phase 3: decreasing loss and hierarchical clustering. Note that due to the sheer size of the AMap dataset, all visualizations were plotted after random sampling of the data instances for practicality.}
    \label{fig:train_dyna}
\end{figure*}

Specifically, we consider the following following variants of SimCIT:
% Specifically, we consider the following variants affect the effectiveness of SimCIT: 
(1) w/o p.h. without the projection head instead applying contrastive loss on the representation $z$ directly. (2) w/o g.s. without the Gumbel search strategy instead using a hard $\arg\min$ search strategy. (3) w/o a.s. without annealing schedule on the temperature $\alpha$ instead using a constant. (4) w/o m.f. without fusing multi-modal information into the identifier learning instead only using the text descriptions like other common settings. 
% (3) We utilize semantic, collaborative, and
% diversity regularization for item tokenization and train TIGER with
% original generation loss, denoted as “(1) w/ d. r.”. (4) We employ
% all regularizations for item tokenization and apply ranking-guided
% generation loss, i.e., LETTER-TIGER.

\begin{figure}
    \centering
        \label{fig:composition}
    \includegraphics[width=0.9\linewidth]{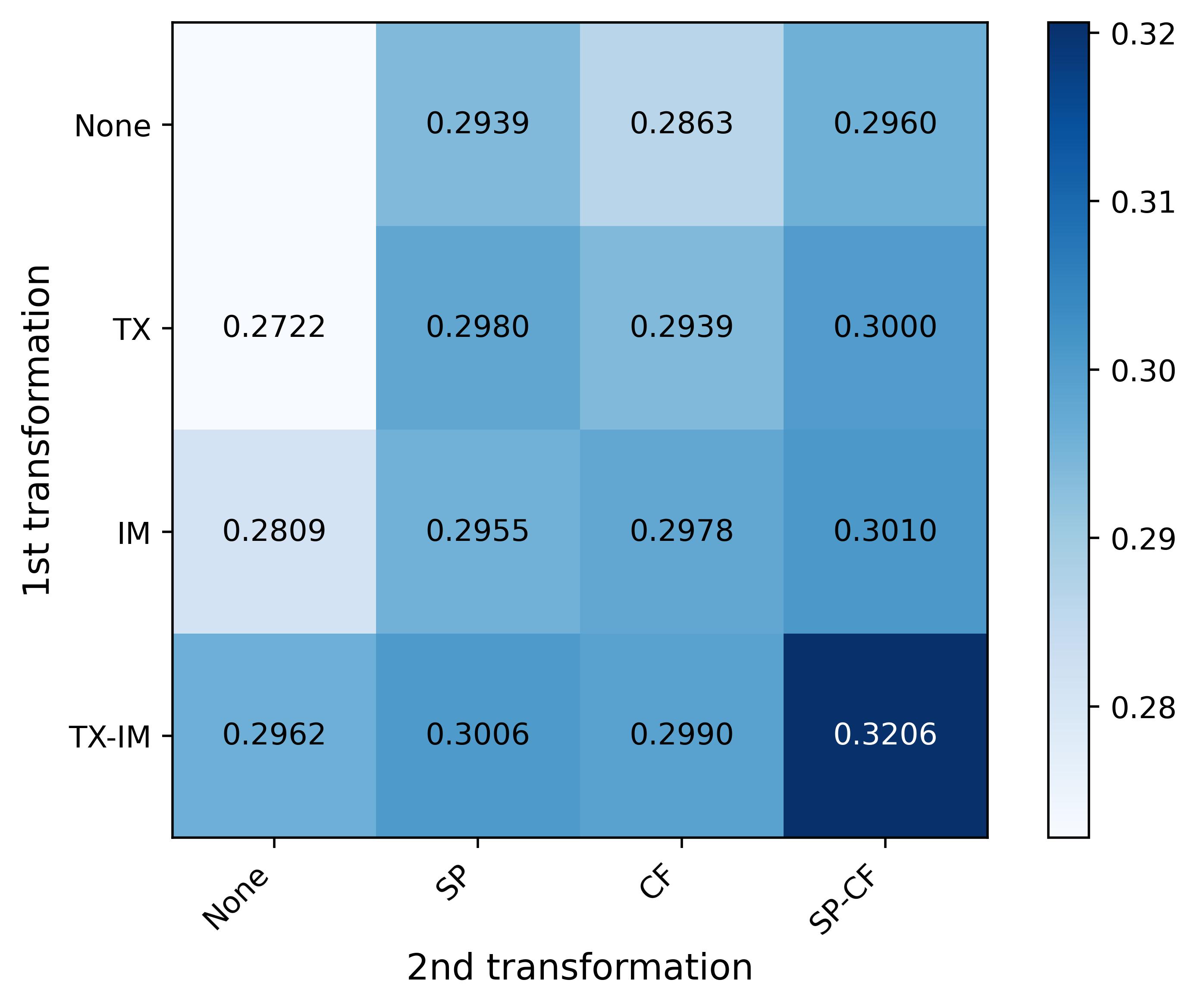}
    \caption{Recall@10 results of AMap under individual or composition of input modals. text is denoted as TX, IM is image, SP is spatial information and CF is collaborative signals. TX-IM (SP-CF) means composition of text-image(spatial-collaborative).}
\end{figure}

\begin{figure*}[h]
    \centering
    \includegraphics[width=\linewidth]{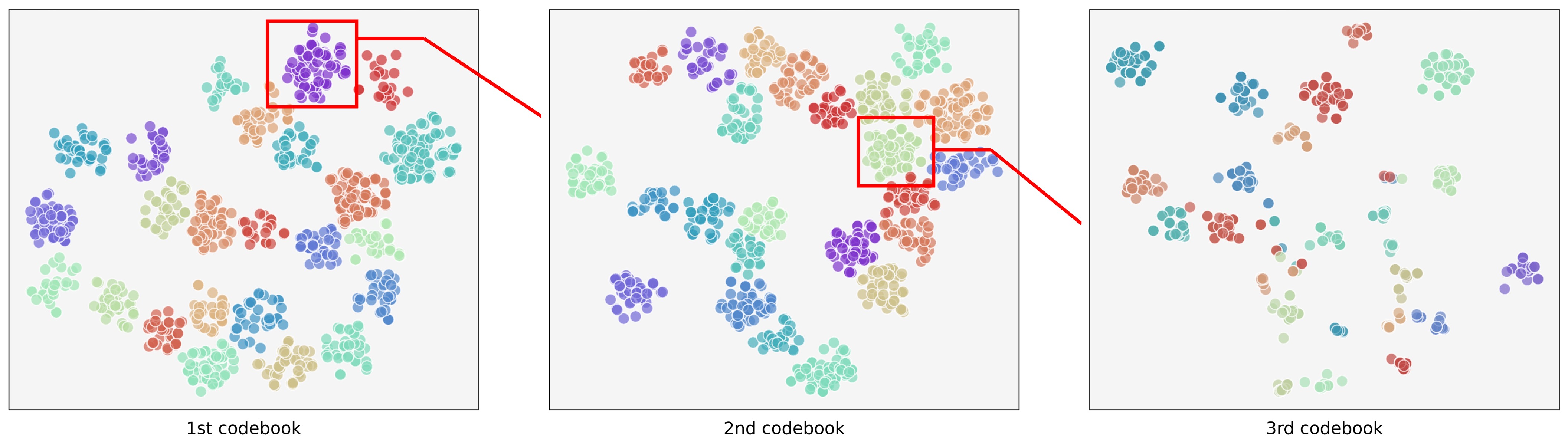}
    \caption{Code embedding distributions results of the 3-Layer codebooks (24 codewords in each book). Each codeword is with a different color. The code embeddings present a ideal hierarchical clustering.
    Note that due to the sheer size of the AMap dataset, all visualizations were plotted after random sampling of the data instances for practicality.}
    \label{fig:hierclu}
\end{figure*}

The experimental results for our SimCIT and its multiple variants are presented in Table \ref{tab:ablation}. As observed, removing any of the aforementioned techniques leads to a decline in overall performance. 
The absence of projection head in computing contrastive loss results in a significant performance decline, as observed in \cite{Chen2020SimCLR,gupta2022understandingimprovingroleprojection}.

The annealing schedule for the Gumbel-Softmax temperature $\alpha$ is imperative — failure to implement this progressive softening ($\alpha: 0.2 \xrightarrow{} 0$) results in codebook collapse and subsequent performance degradation. Hard search substitution catastrophically disrupts gradient flow, causing precipitous effectiveness drops(variants (2) and variants (3)).
Removing multi-modal semantic information (i.e., variants (4)) results in a lack of collaborative and spatial knowledge within the item tokenizer, causing performance degradation. The synergistic effects of different multi-modal compositions, particularly their differential impacts on recommendation robustness and semantic alignment, will be systematically investigated in Section \ref{further_analysis}.
% applying the pre-trained item tokenizer to downstream datasets (i.e.,
% variant (4)) fails to achieve the expected results, which is likely due
% to the entanglement of item codes in the new domain. Furthermore,
% fine-tuning all codebook parameters (i.e., variant (5)) may disrupt
% the associations between codes, resulting in the loss of general
% knowledge acquired during pre-training.

\subsection{Further Analysis}
\label{further_analysis}

\subsubsection{Multi-modal Synergy Analysis}
SimCIT demonstrates native compatibility with heterogeneous item information sources through unified identifier alignment. We systematically evaluate modality-specific contributions using AMap containing four distinct modalities: (1). textual descriptions (denoted as TX), (2). image content (denoted as IM), (3). Spatial information (denoted as SP) and (4). collaborative filtering signals (denoted as CF).
As shown in Fig. \ref{fig:composition}, progressive modality integration yields monotonically improving performance in generative recommendation tasks. Crucially, spatial features provide the largest individual gain. The full quad-modal configuration achieves state-of-the-art results, validating our framework's capacity to synthesize cross-modal synergies and preserve modality specificity.

\subsubsection{Training Dynamics Analysis}
In this section, we experimentally demonstrate SimCIT's training process along with the associated learning dynamics of the identifier. 
Throughout the entire training process, as illustrated in Fig. \ref{fig:train_dyna}, 
% which allow us to divide the entire training process into 
three phases based on the variation of loss and perplexity are observed.

% \begin{figure}[h]
%     \centering
%     \includegraphics[width=\linewidth]{figs/scatters.png}
%     \caption{Training dynamics of the loss and perplexity of the identifier on Foursquare NYC dataset.}
%     \label{fig:train_dyna}
% \end{figure}
In the initial phase of training, 
we observe steep loss decay concurrent with identifier perplexity stagnation, indicating predominant modality-level discrimination. During this regime, the model establishes coarse cluster boundaries between reconstructed samples and naive cross-modal negatives, yet suffers high inter-identifier collision rates. 

Subsequently, 
cluster dispersion initiates as perplexity escalates, but inversely correlates with hard negative discrimination capacity deterioration. This entropy-quality duality manifests as non-monotonic loss progression, suggesting the competing optimization objectives between semantic preservation and identifier diversity.

Finally, the system converges to Pareto-optimal equilibrium where gradients exhibit directional consensus toward joint optimization of multi-modal fusion and collision-aware diversity regularization.
% the loss decreases rapidly, whereas the identifier's perplexity experiences little to no change. During this stage, the model merely distinguishes between reconstructed samples and simple samples of other modalities, but a large number of reconstructed samples are clustered together, resulting in a high collision rate.
% Subsequently, the reconstructed samples start to disperse, and the identifier's perplexity gradually increases. However, during this phase, the model's ability to distinguish simple negative samples declines, resulting in a rise in loss.
% 3. In the final stage, the model finds the optimal direction that leads to a decrease in loss and continues to optimize towards integrating multi-modal semantics and identifier diversity.
This validates the effectiveness of diversity regularization to achieve a more diverse distribution of code embeddings in the representation space, fundamentally alleviating the problem of biased code assignment.

\begin{figure*}
    \centering
    \includegraphics[width=\linewidth]{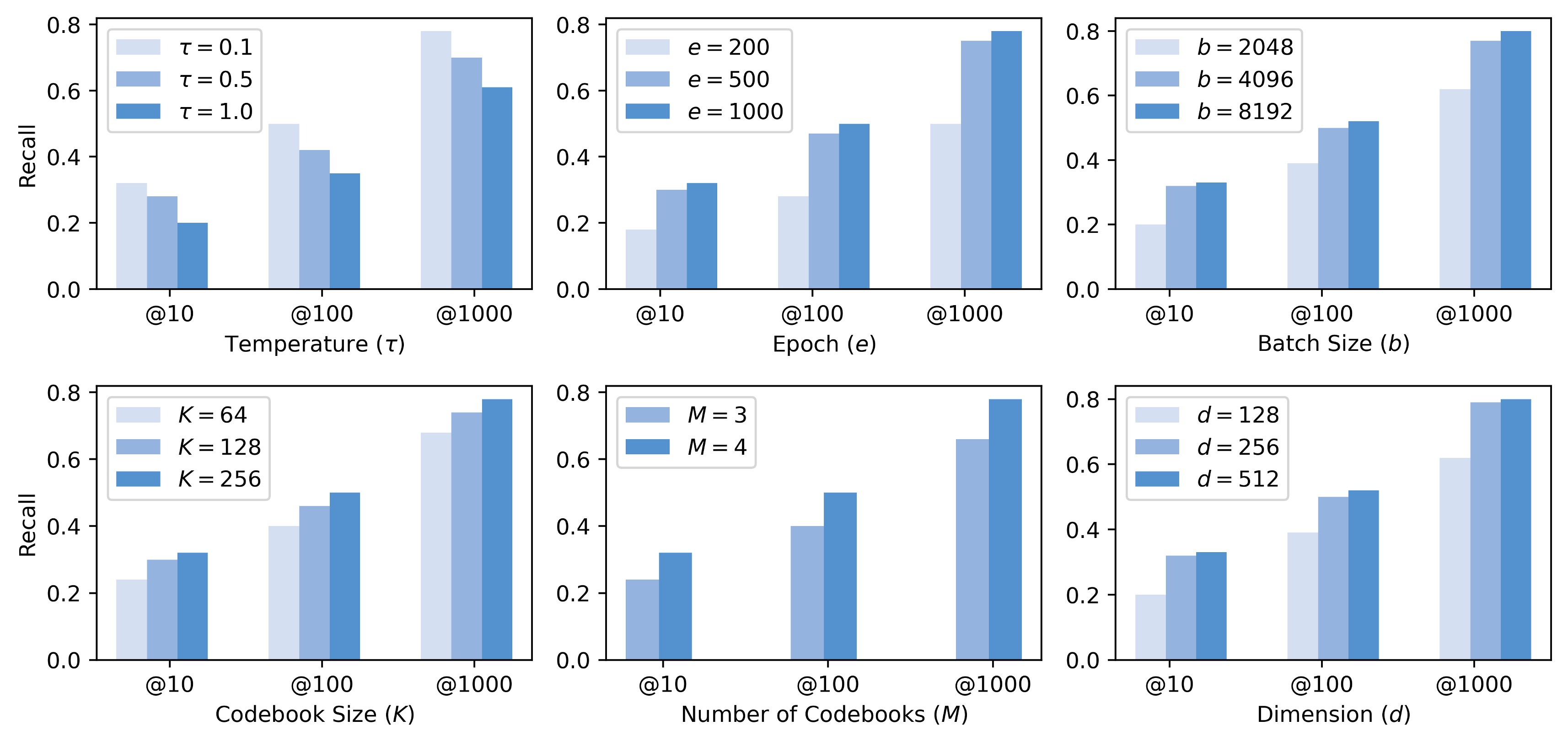}
    \caption{Ablation study of (top) temperature($\tau$), epoch($\textit{e}$), batch size ($\textit{b}$) and (bottom) codebook size $K$, number of codebooks $M$, embedding dimension $d$ on the Amap dataset.}
    \label{fig:abla}
\end{figure*}

\subsubsection{Code Assignment Distribution}
We further investigate whether the contrastive loss enforce an discriminative regularization and effectively mitigate the code assignment bias in item tokenization. 
In this section, we conduct empirical validation on the AMap dataset employing 3-level codebooks for SimCIT, where each codebook comprises 256 code embeddings with a dimension of 24.
To facilitate interpretation, we project the code embeddings into 2-dimensional space via t-SNE. Hierarchical codebook visualizations (Fig. \ref{fig:hierclu}) depict code embeddings as chromatic-encoded nodes. 
In the first codebook, Fig. \ref{fig:hierclu} (left)
demonstrates uniform categorical distribution (24 classes), where each cluster bijectively maps to discrete codebook identifiers.
Fig. \ref{fig:hierclu} (center) present one cluster of the 1st codebook.
The second codebook layer preserves this partitioning fidelity - when visualizing embeddings associated with any single 1st identifier, they further stratify into 24 distinct 2nd sub-clusters. 
The 3rd layer (Fig. \ref{fig:hierclu} (right)) maintains equivalent differentiation capacity.
This sustained hierarchical differentiation evidences SimCIT's effectiveness in learning multi-granular item taxonomy, which reduces generative retrieval computational complexity and improve the performance of generative recommendation.

\subsubsection{Sensitivity Analysis of Temperature $\tau$, Batch Size and Training Epochs}
To analyze how the temperature $\tau$ affects the training of SimCIT, we conducted experiments on the Amap dataset with $\tau$ set to \{0.1, 0.5, 1.0\}. SimCIT was trained 1000 epochs with a batch size of 8192, yielding the best Recall@10 when $\tau$ = 0.1, as shown in Fig. \ref{fig:abla} (top). This indicates that the performance of SimCIT varies with $\tau$ within a specific range. 
Additionally, fixing $\tau$ = 0.1, we conducted extensive experiments with progressively increasing batch sizes (2048, 4096, 8192) across 1000 epochs. The observed performance improvement showed strong positive correlation with batch size magnitude, empirically validating the enhanced efficacy of larger batches in contrastive learning frameworks.

Lastly, fixing $\tau$ = 0.1 and batch size = 8192, we trained the model for 200, 500, and 1000 epochs. As illustrated, with an increase in the number of training epochs, the loss steadily decreases, and the Recall metrics consistently improve, demonstrating the effectiveness of SimCIT in training generative recommendation models.

\subsubsection{Sensitivity Analysis of Codebooks} The effectiveness of semantic tokenization can be influenced by key hyperparameters, such as the codebook size ($K$) the number of codebooks ($M$), and the embedding dimension ($d$) of code vectors. As illustrated in Fig. \ref{fig:abla} (Bottom), as the codebook size and the number of codebooks increases, NDCG
shows a consistent upward trend. This suggests that a larger token space allows for more effective representation of each item, leading to more precise tokenization. Additionally, as the embedding dimension increases from 128 to 512, the model’s capability to represent items improves, leading to enhancements across various metrics. However, increasing the dimension to 256 results in a slight performance degradation due to overfitting.

\section{Conclusion}
\label{sec:conclusion}
In this work, we delve deeper into semantic tokenization as a crucial preliminary step for generative recommendation tasks. To overcome the limitations of reconstruction quantization-based RQ-VAE, we introduce a novel contrastive quantization framework into the tokenization process, resulting in the SimCIT method. By leveraging contrastive loss, which maximizes the top-one probability within a batch, SimCIT more effectively captures the underlying distribution of item similarities, providing a key advantage for recommendation tasks. Our experimental results demonstrate the effectiveness of SimCIT in enhancing generative recommendation performance. While SimCIT currently focuses on top-one neighborhood alignment, future work could extend this to consider the generative recommendation framework. Looking ahead, we anticipate that further advancements in semantic tokenization, potentially aligning the natural language token space, will continue to improve the efficacy of generative recommender systems.

\section*{Acknowledgment}

We would like to acknowledge the discussions of Lingyue Zhong, Junchi Xing, and Fangfang Chen throughout this work. Furthermore, we are grateful for Tucheng Lin, Jian Song, Tingting Hu, Shulong Han, Jinhui Chen, Xiongfei Fan, Zudan Cao, Jing sun, Yifei Fan and Yukun Liu, for their technical and engineering assistance.

\bibliography{reference.bib}

\end{document}